\documentclass[a4paper,12pt]{article}
\usepackage{a4wide}
\usepackage{graphicx}
\usepackage{amsmath}
\usepackage{amssymb}
\usepackage[latin1]{inputenc}
\usepackage[english]{babel}
\usepackage{url}
\usepackage{cite}

\title{PYBBWH: A program for associated charged Higgs and $W$ boson 
production}
\author{David~Eriksson\\[4mm]
\normalsize
High Energy Physics, Uppsala University,
Box 535, S-75121 Uppsala, Sweden\\
\normalsize
E-mail: \texttt{david.eriksson@physics.uu.se}
}
\date{February 3, 2009}

\begin{document}

\maketitle

\begin{abstract}
The Monte Carlo program, PYBBWH, is an implementation of the associated 
production of a charged Higgs and a $W$ boson from $b\bar b$ fusion in 
a general Two-Higgs-Doublet model for both CP-conserving and CP-violating 
couplings.
It is implemented as a external process to \textsc{Pythia 6}.
The code can be downloaded from \texttt{http://www.isv.uu.se/thep/MC/pybbwh}.
\end{abstract}

\section{\boldmath Associated $H^\pm$ and $W$ boson production}

\begin{figure}
\centering
\begin{tabular}{c@{\hspace{1cm}}c}
\includegraphics{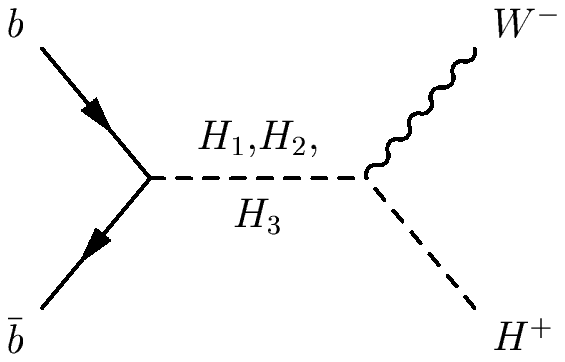} &
\includegraphics{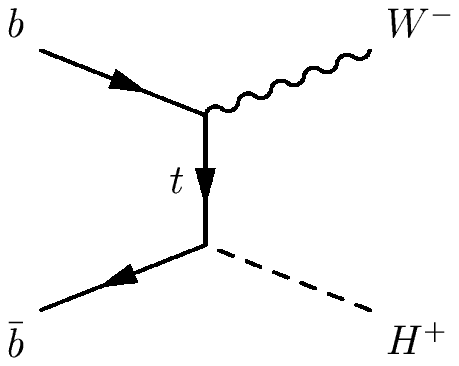}
\end{tabular}
\caption{Feynman diagrams for $H^\pm W^\mp$ production via
$b\bar{b}$ annihilation, $b \bar b \rightarrow H^+ W^-$. }\label{feynsignal}
\end{figure}

The program, PYBBWH, is an implementation of the production of a charged Higgs
boson, $H^\pm$, in association with a $W$ boson. The code was developed 
for the research presented in \cite{Eriksson:2006yt} where details on the 
theory and numerical results are presented.
This program is written for a general Two-Higgs-Doublet model type II. 
The dominant production mode at tree level occurs is via $b\bar b$ fusion and 
at one-loop-level via gluon fusion. 
This program implements the leading order $b\bar b$ fusion part and the 
relevant Feynman diagrams are given in figure \ref{feynsignal}. By only 
including $b\bar b$ fusion this program is most suited for intermediate 
$H^\pm$ masses and large $\tan \beta$.

In a general type II 2HDM the couplings relevant for this production can be 
specified via the Higgs mixing matrix $O_{ji}$ in the following way
\begin{equation}\label{couplings}
\begin{aligned}
g_{H_i H^- W^+}&= g_{H_i H^+ W^-}^* = 
  O_{2i}\cos\beta-O_{1i}\sin\beta+ \mathrm{i} \, O_{3i} \,,\\
g_{H_i \bar{b} b}&=O_{1i}+ \mathrm{i} \, O_{3i}\sin\beta \,.
\end{aligned}
\end{equation}
In a real 2HDM were $H_i=\{h^0,H^0,A^0\}$ the mixing matrix has the simple form
\begin{equation}
O_{ji}=
\begin{pmatrix}
-\sin\alpha & \cos\alpha & 0 \\
\cos\alpha & \sin\alpha & 0 \\
0 & 0 & 1 \\
\end{pmatrix}
\end{equation}
which gives purely real couplings for $h^0$, $H^0$ and imaginary couplings for 
$A^0$.
In a general 2HDM there can be mixing between the CP-even and CP-odd Higgs 
states and the mixing matrix can have all elements non-zero. 

Using a formalism with only diagonal propagators for the Higgs bosons\footnote{This formalism is valid as long as the Higgs bosons are well separated in mass. If they are so close that they overlap non-diagonal propagators are needed and this program can not be used.}, the 
differential cross-sections implemented in this program for the two processes 
are~\cite{BarrientosBendezu:1998gd,Akeroyd:2000zs}: 
\begin{multline}
\frac{d\sigma}{dt}(b \bar b \rightarrow H^+ W^-)=
\frac{G_F^2}{ 24\pi s}\\
\left\{\frac{m_b^2 \lambda(s,m_W^2,m_{H^\pm}^2)}{2\cos^2\beta}
\sum_{i,j}g_{H_i H^- W^+}g_{H_j H^- W^+}^*S_{H_i}S_{H_j}^*
\mathrm{Re}[g_{H_i \bar b b}g_{H_j \bar b b}^*]\right.\\
+\frac{1}{(t-m_t^2)^2}\left[m_t^4 \cot^2\beta (2m_W^2+p_\perp^2)
+m_b^2 \tan^2\beta(2m_W^2p_\perp^2+t^2)\right]\\
\left.+\frac{m_b^2 \tan\beta}{(t-m_t^2)\cos\beta}\left[m_W^2 m_{H^\pm}^2-s p_\perp^2-t^2\right]
\sum_{i}\mathrm{Re}\left[g_{H_i H^- W^+}g_{H_i \bar b
    b}S_{H_i}\right]\right\} ,
\label{eq:bbhpwm}
\end{multline}
\begin{multline}
\frac{d\sigma}{dt}(b \bar b \rightarrow H^- W^+)=
\frac{G_F^2}{24\pi s}\\
\left\{\frac{m_b^2 \lambda(s,m_W^2,m_{H^\pm}^2)}{2\cos^2\beta}
\sum_{i,j}g_{H_i H^- W^+}^*g_{H_j H^- W^+}S_{H_i}S_{H_j}^*
\mathrm{Re}[g_{H_i \bar b b}^*g_{H_j \bar b b}]\right.\\
+\frac{1}{(t-m_t^2)^2}\left[m_t^4 \cot^2\beta (2m_W^2+p_\perp^2)
+m_b^2 \tan^2\beta(2m_W^2p_\perp^2+t^2)\right]\\
\left.+\frac{m_b^2 \tan\beta}{(t-m_t^2)\cos\beta}\left[m_W^2 m_{H^\pm}^2-s p_\perp^2-t^2\right]
\sum_{i}\mathrm{Re}\left[g_{H_i H^- W^+}^*g_{H_i \bar b
    b}^*S_{H_i}\right]\right\} ,
\label{eq:bbhmwp}
\end{multline}
where $s$ and $t$ are the ordinary Mandelstam variables of the hard process and
\begin{equation}
\lambda(x,y,z)=x^2+y^2+z^2-2(xy+yz+zx) \, ,
\end{equation}
\begin{equation}
p_\perp^2 = \frac{\lambda(s,m_W^2,m_{H^\pm}^2) \sin^2\theta}{4s}  \, ,
\end{equation}
with $\theta$ being the polar angle in the $2\to2$ cms and
\begin{equation}
S_{H_i}=\frac{1}{s - m_{H_i}^2 + i m_{H_i} \Gamma_{H_i}}
\label{eq:SHi}
\end{equation}
the propagators of the neutral Higgs bosons.

\section{Implementation}

The associated production program is implemented as an external process to 
\textsc{Pythia 6}. It uses the Les Houches generic user process interface
for event generators \cite{Boos:2001cv} but it also uses \textsc{Pythia} 
specific routines. This means it can not be used directly together with other
event generators, but after some minor changes to the code it should be 
possible.
The code has been tested with with \textsc{Pythia} version 6.324 and the most 
recent version 6.413. Details on how to use an external process can be found in
the \textsc{Pythia 6} manual \cite{Sjostrand:2006za}, section 9.9. In principle
this is done by initiating \textsc{Pythia} with 
\texttt{PYINIT('USER',' ',' ',0d0)}. The program  does not work directly with 
\textsc{Pythia 8} but it can be made to work via the Les Houches interface in 
\textsc{Pythia 8}.

The widths of the $H^\pm$ and $W$ bosons are included in the same way as in
standard \textsc{Pythia} whenever possible, see details below. In other words 
the $H^\pm$ and $W$ masses vary according to Breit-Wigner distributions
with mass dependent widths meaning that for each mass the decay widths are 
recalculated based on the open decay channels.

The program is setup for simulating LHC events as default, meaning incoming 
protons with 7 TeV energy. This can be changed by setting the parameters
\texttt{IDBMUP} and \texttt{EBMUP} to incoming particle type and energy 
respectively. These parameters has to be set before \texttt{PYINIT} is called.

\subsection{{\normalfont\scshape Pythia} specified couplings}

\textsc{Pythia} contains different SUSY simulations. The default in the 
PYBBWH program is to assume that one of these is used, so that the Higgs 
mixing angle $\alpha$ and $\tan \beta$ are given in \texttt{RMSS(18)} and 
\texttt{RMSS(5)}, respectively. The Higgs mixing matrix and the couplings 
used in the process generation is then calculated from $\alpha$ and 
$\tan \beta$. Some checks are performed to see if the values of 
\texttt{RMSS(18)} and \texttt{RMSS(5)} correspond to the Higgs-fermion 
couplings used in \textsc{Pythia} for Higgs decay. A warning is printed if an 
inconsistency is detected.

\subsection{User specified couplings}

If an external SUSY simulation or a general 2HDM type II is used, especially 
one with CP-violation, the default method can no longer be used. In this case 
the common block PYBBWH is used. It is defined as
\begin{verbatim}
      INTEGER IPYBBWH
      DOUBLE PRECISION O_M
      COMMON/PYBBWH/O_M(3,3),IPYBBWH
\end{verbatim}
where \verb!O_M(3,3)! is the Higgs mixing matrix. \verb!IPYBBWH! is a switch 
with the default value of \verb!0!, meaning \textsc{Pythia} specified 
couplings. For \verb!IPYBBWH=1!, \verb!O_M(3,3)! is used to calculate the 
couplings. Also in this case $\tan \beta$ is taken from \texttt{RMSS(5)}.
The masses and widths for all Higgs particles must also be set correctly in 
the \verb!PMAS! array.

Another difference between \verb!IPYBBWH=0! and \verb!IPYBBWH=1! is that for 
the former a varying width is used for the charged Higgs but for the later a 
fixed width, the one given by \verb!PMAS(37,2)!, is used. This behavior is 
motivated since if the Higgs mixing matrix is specified  in \verb!O_M! the 
couplings used in \textsc{Pythia} to calculate the Higgs decay are probably 
wrong. If ones wants to use a varying width in this case this can be done by 
substituting the function 
\begin{verbatim}
      DOUBLE PRECISION FUNCTION HPWID(MHP)
\end{verbatim}
with a new function that gives the correctly varying width. 
The argument given to \verb!HPWID! is defined as
\begin{verbatim}
      DOUBLE PRECISION MHP
\end{verbatim}
and is the mass at which the width is to be calculated.

\section{Download}

The code can be downloaded from \texttt{http://www.isv.uu.se/thep/MC/pybbwh}.
On that web page there is also two example programs, one for {\normalfont\scshape Pythia} specified couplings and one for User specified couplings.

\section{Final comments}

This code was developed for research published in \cite{Eriksson:2006yt}. The 
work was done in collaboration with Stefan Hesselbach and Johan Rathsman. 
If you use the code please cite \cite{Eriksson:2006yt} and this manual.

\end{document}